\documentclass[conference]{IEEEtran}
\ifCLASSINFOpdf
  \usepackage[pdftex]{graphicx}
  \graphicspath{{./img/}}
\else
  \usepackage[dvips]{graphicx}
  \graphicspath{{./img/}}
\fi
\begin{document}
\title{Distributed Lustre activity tracking}

\author{\IEEEauthorblockN{Henri Doreau - CEA/DAM}
\IEEEauthorblockA{CEA, DAM, DIF, F-91297 Arpajon, France\\
Email: henri.doreau@cea.fr}}

\maketitle

\begin{abstract}
%\boldmath
Numerous administration tools and techniques require near real time vision of
the activity occuring on a distributed filesystem. The changelog facility provided
by Lustre to address this need suffers limitations in terms of scalability and
flexibility. We have been working on reducing those limitations by enhancing
Lustre itself and developing external tools such as \emph{Lustre ChangeLog
Aggregate and Publish (LCAP)} proxy. Beyond the ability to distribute changelog
processing, this effort aims at opening new prospectives by making the changelog
stream simpler to leverage for various purposes.
\end{abstract}

\section{Introduction}
The Lustre filesystem provides external tools with activity tracking facility.
It is implemented as a stream of records, describing every modifying metadata
operation. Administrators have to enable this feature and manually register each
reader. Those readers then asynchronously poll the metadata servers for new
records in order to follow the filesystem activity.

Though already proven valuable for tools like \emph{Robinhood Policy Engine},
this mechanism had to be modified for its usages to be further expanded.
Robinhood reads changelogs to replicate filesystem changes into a database and
take decisions based on the observed events \cite{Rbh}.

The first encountered issue was scalability: Robinhood reads changelogs to
update its database near real time, but with features like \emph{Distributed
NamespacE (DNE)} \cite{DNE} and more generally with the increasing ability of
Lustre to scale, it will be necessary to distribute the processing of the
changelog stream.  Preferably, this could be done without assigning an instance
of robinhood per MDT \cite{LustreComponents} but in a load-balanced fashion. Our
tool LCAP (for \emph{Lustre Changelog Aggregate and Publish} is a proxy that
aims at doing so.

A second area of concern is the ability of the changelog stack to handle
emerging usages. We have been focusing on two aspects: enriching the records, by
introducing a compatible and extensible format, and supporting new kinds of
changelog consumers, such as ephemeral readers. This work will be described in
the final section of this paper.

% Section I
% Comment ca fonctionne??!
\section{Existing infrastructure}
Changelogs were introduced with Lustre 2.0. To leverage the feature, an
administrator must manually enable it, then register readers manually. As soon
as there is a registered reader, metadata operations are logged in a persistent
journal. The administrator can select which operations to log. The records are
kept on MDT disks, until read and acknowledged by all registered
readers \cite{ChangelogsAPI}

Changelog records are exported to userland programs running on Lustre clients.
They have to do a four phase loop. First register to a given MDT, to open a
communication channel and indicate the index of the first desired record.  Then
receive a single record. Once processed, the record has to be acknowledeged
(this can be delayed and batched). For this operation the process identifies
itself with its reader ID. Finally, on demand or after having reached the end of
stream, the process closes the pipe, to possibly start again.

\begin{itemize}
 \item Register
 \begin{enumerate}
 \item Start
  \item{Receive/Consume/Free record}
  \item{Acknowledge consumed record(s)}
  \item Stop
  \end{enumerate}
 \item Deregister
\end{itemize}

Two limitations of Lustre changelogs are visible: first, the start command is
not issued for a given reader ID but for a changelog index on a given MDT.
Second, registration of a new reader has to be made manually, server-side.

The changelog storage and distribution stack mainly relies on two internal
Lustre mechanisms, namely \emph{Lustre log} (LLOG), and \emph{kernel-userland
communication} (KUC). Both subsystems offer room for improvement in term of
efficiency and flexibility.

% Section II
% Distributed processing, rbh, lcap
\section{Distributed changelog processing}
Since version 2.4, Lustre offers the ability to work with several metadata
servers. A typical robinhood configuration consists of a single instance, with
one thread per MDT and a centralized database. This will eventually be unable
to keep up with the pace of changes occuring in the namespace. To address this,
we are working on distributing its database and distributing the processing of
the changelog stream. This document focuses on the second aspect.

\subsection{LCAP}
Assuming we have multiple instances of robinhood operating on a shared database,
we want the changelog streams originating from the MDTs to be aggregated and
spread evenly among the instances.

Our approach is to develop a changelog proxy, called LCAP (for \emph{Lustre
Changelog Aggregate and Publish}) which behaves like a regular changelog reader
but maintains lists of consumers and redistribute the records to them in
versatile ways.

LCAP proxy uses a client/server architecture to aggregate, pre-process and
redistribute Lustre changelog records. Its goal is to act as a broker between
one or multiple producers (the Lustre MDTs) and consumers (like instances of
robinhood). The system must deal with a potential imbalance between the number
of actors and variable rates of the incoming and outgoing streams: burst of
records emitted, sudden slowdown of a consumer\ldots Thus LCAP has to be a load
balancer.

It introduces the concept of consumer group. Multiple processes start reading
changelogs and identify themselves as members of a same group.  The stream is
spread among instances of a single group, thus load-balancing record processing.
If multiple groups co-exist, every record will be delivered to each group.
Records will be acknowledged upstream to Lustre only once acknowedged by every
group. This means we use a \emph{at least once} delivery strategy.

LCAP adopts a \emph{greedy} behavior, to read records as soon as possible and load
them into memory. Persistence of records is left to Lustre, since we accept an
\emph{at least once} delivery strategy. Along with batching, these aspects are
crucial in LCAP performances.

LCAP is written in C, and distributed under the terms of the GNU LGPLv3+
license. It is fully lockless, and uses the ZeroMQ library for internal and
external communications \cite{ZMQLib}. The server relies on modules, implemented
as shared libraries, to pre-process the stream as desired. For instance, records
can be dropped for operations that compensate each others (creat/unlink) or
re-ordered to optimize downchain processing.

\begin{figure}[!t]
 \centering
 \includegraphics[width=3.5in]{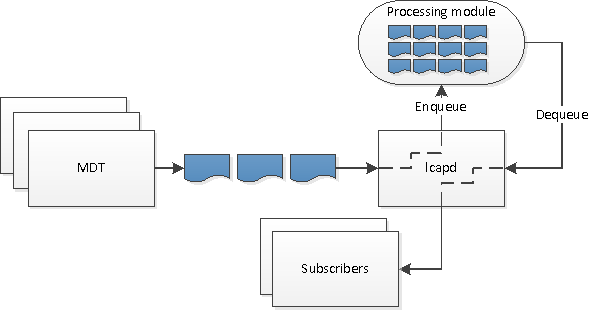}
 \caption{LCAP high level architecture}
 \label{fig_arch}
\end{figure}

\begin{figure}[!t]
 \centering
 \includegraphics[width=3.5in]{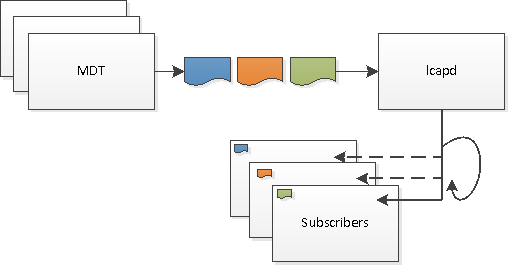}
 \caption{Load balanced processing with LCAP}
 \label{fig_loadbl}
\end{figure}

% Section III
% Format
\section{Towards new usages of changelogs}
Though it was proven robust for some uses, the current changelog reader
registration model is very rigid and thus not well suited for other
applications. With new usages come new requirements, and we have identified the
need to add information to the emitted changelog records. Second, we wanted to
relax the constraints related to registration and more generally, make it easy
to implement new changelog distribution patterns.

\subsection{Extensible changelog format}
While exploring the use of Lustre changelogs as a notification mechanism for
other tools working on top of Lustre, we faced the need to enhance the records
and make them carry other information. Typically, the JobID of the process from
which the described operation originated. Such an enhancement had already been
done (as of LU-1331) to emit a single \emph{extended} record after a rename but
it was neither fully compatible with existing applications nor easy to further
extend due to the chosen approach of a second data structure.

The work done along with intel as of LU-1996 introduces flexibility,
demonstrated by the addition of a new jobid field in the records, while
preserving compatibility between newer and older versions of applications,
clients and servers. It introduces the required infrastructure to add new fields
to the changelog records in an easy and compatible way.

\begin{figure}[!t]
 \centering
 \includegraphics[width=3.5in]{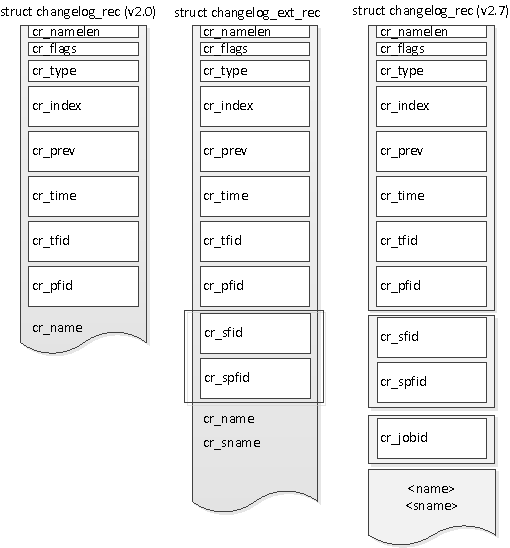}
 \caption{Changelog structures, before and after}
 \label{fig_struct}
\end{figure}

Changelog consumers express through flags the list of extra fields that they
need, and records get remapped to fit this format. Remapping happens locally or
remotely, depending on whether fields are to be added (expected by a recent
client but not available on the server) or removed (not expected by an older
client but present on the server). This work has been made available as of
Lustre 2.7.

Extensions and final variable-length fields are accessed via inline functions
which compute the right offsets according to the structure format as described
by flags. The figure \ref{fig_struct} illustrates the two older record formats
along with the new one we introduced.

Within lustre, where no assumptions are being made on the format of the records,
it is already possible to leverage this to manipulate heterogeneous records
efficiently spending neither disk space nor bandwidth to store/send oversized
records with empty fields.

\subsection{Versatile distribution patterns}
The existing changelog infrastructure enforces strict distribution scheme. We
want to be able to loosen that to adapt it to our needs.

First comes the need for ephemeral changelog consumers. Reader which would only
be interested in records emitted during their execution time. In other words,
consumers that do not want to receive anything that has happened before they
started or that could happen after they have stopped (in case of a restart).
Similar to a listener following a radio broadcast. This is difficult to
implement efficiently using only the liblustreapi infrastructure.

Then comes the need for distributed workers processing the changelog stream
collaboratively. The changelog reading API as exposed by lustre requires
consumers to indicate at which offset of the llog they want to start, then
dequeue records sequentially until EOF. This is not well-suited to distributed
processing. Especially in the case of Robinhood, where proper load balancing
would make sense. In this particular case, and unlike the one above, not loosing
any record is fundamental.

LCAP proxy implements such distribution patterns. Actual changelog readers (LCAP
clients) can identify themselve as ephemeral or persistent. Ephemeral readers do
not receive records emitted before their connection to the proxy and they are
not expected to acknowledge records.  Persistent readers receive everything and
acknowledgement is made upstream based on what they have actually acknowledged
collectively.

\subsection{Usage examples}
\subsubsection{pNFS servers}
The userland NFS/9P server \emph{Ganesha} \cite{NFSGanesha} can act as a LCAP
client to receive changelogs and get notified of what other instances (based on
the JOBID field) did on the filesystem. Such notifications are a NFSv4.1
requirement as stated by RFC 5661 \cite{RFC5661}. Ganesha uses Lustre changelogs
as a loose metadata cache invalidation mechanism. Thanks to LCAP allowing
ephemeral readers, I/O proxies can be spawned on demand at a very low price.
This forms the core of a high efficiency I/O delegation system on top of Lustre.

\vfill
\break

\subsubsection{Fast lustre object index traversal}
Regular POSIX scans such as the ones used to initially populate robinhood
database become difficult to run against filesystems of hundreds of millions of
inodes or more. We are considering the use of a special changelog stream, filled
with entries from the MDT object index, and consumed by instances of the
policy engine. We believe that this could significantly improve the scanning
time. This might require new extensions of the record format, and would
certainly benefit from being consumed by multiple LCAP clients.

\section{Conclusion}
With the increase of scalability of Lustre, up to billions of entries, being
able to efficiently and reliably report the filesystem activity to external
tools is becoming critical for monitoring, administrative or applicative
purposes. The existing changelog infrastructure is undoubtly a sturdy basis for
such tasks but it suffers limitations.  We propose a changelog proxy, called
LCAP, to help designing userland solutions relying on changelogs, with relaxed
conditions where needed, and improved scalability. Additionally, we have made
changes to the changelog distribution stack so that records are easily
extensible over time and versions of Lustre. These efforts, combined, allow us
to envision new fields of applications for changelogs.

%% trigger a \newpage just before the given reference
%% number - used to balance the columns on the last page
%% adjust value as needed - may need to be readjusted if
%% the document is modified later
%%\IEEEtriggeratref{8}
%% The "triggered" command can be changed if desired:
%%\IEEEtriggercmd{\enlargethispage{-5in}}
%
%% references section
%
%% can use a bibliography generated by BibTeX as a .bbl file
%% BibTeX documentation can be easily obtained at:
%% http://www.ctan.org/tex-archive/biblio/bibtex/contrib/doc/
%% The IEEEtran BibTeX style support page is at:
%% http://www.michaelshell.org/tex/ieeetran/bibtex/
\bibliographystyle{IEEEtran}
%% argument is your BibTeX string definitions and bibliography database(s)
\bibliography{IEEEabrv,./bibliography}
%%
%% <OR> manually copy in the resultant .bbl file
%% set second argument of \begin to the number of references
%% (used to reserve space for the reference number labels box)
%\begin{thebibliography}{1}
%\end{thebibliography}

\end{document}